\documentstyle[prl,aps,floats,epsfig]{revtex}
\tighten
\let\jnfont=\rm
\def\NPB#1,{{\jnfont Nucl.\ Phys.\ }{\bf B#1},}
\def\PLB#1,{{\jnfont Phys.\ Lett.\ B }{\bf #1},}
\def\PRD#1,{{\jnfont Phys.\ Rev.\ D }{\bf #1},}
\def\PRL#1,{{\jnfont Phys.\ Rev.\ Lett.\ }{\bf #1},}
\def\ZPC#1,{{\jnfont Z.~Phys.\ C }{\bf #1},}

\def\gsim{\mathrel{\mathpalette\oversim>}}
\def\oversim#1#2{\lower0.5ex\vbox{\baselineskip0pt\lineskip0pt
  \lineskiplimit0pt\everycr{}\tabskip0pt
  \halign{$\mathsurround0pt #1\hfil##\hfil$\crcr #2\crcr\sim\crcr}}}
\begin{document}
\draft
\preprint{}

\title{ Heavy Supersymmetric Particle Effects in Higgs Boson Production \\ 
        Associated with a Bottom Quark Pair at LHC and Tevatron}
 
\author{\ \\[2mm] Guangping Gao}
\address{\ \\[1mm]
   {\it Institute of Theoretical Physics, Academia Sinica, Beijing 100080, China} }
\author{\ \\[1mm] Robert J. Oakes}
\address{ \ \\[1mm]
   {\it Department of Physics and Astronomy, Northwestern University,
              Evanston, IL 60208, USA} }
\author{\ \\[1mm] Jin Min Yang}
 \address{ \ \\[1mm]
   {\it CCAST(World Laboratory), P.O.Box 8730, Beijing 100080, China;} \\ [1mm]
   {\it Institute of Theoretical Physics, Academia Sinica, Beijing 100080, China}}

\maketitle
\vspace*{0.5cm}

\begin{abstract}
If all the supersymmetry particles (sparticles) except a light Higgs boson are too heavy to be 
directly produced at the Large Hadron Collider (LHC) and Tevatron, a possible way to reveal 
evidence for supersymmetry is through their virtual effects in other processes. We examine such 
supersymmetric QCD effects in bottom pair production associated with a light Higgs boson at the 
LHC and Tevatron.
We find that if the relevant sparticles (gluinos and squarks) are well above the TeV 
scale, too heavy to be directly produced, they can still have sizable virtual effects in this 
process. For large $\tan\beta$, such residual effects can alter the production rate by as much 
as 40 percent, which should  be observable in future measurements of this process.  

\end{abstract}
\pacs{14.80.Ly, 14.80.Cp, 12.60.Jv}

\section{INTRODUCTION}
\label{sec:intro}

Searching for the Higgs boson is one of the most important tasks in particle physics.
The existence of a relatively light Higgs is suggested by high precision fits to the data in the 
Standard Model (SM) and also is theoretically favored in the Minimal Supersymmetric 
Standard Model (MSSM)~\cite{HaberKane}.   Verification  of the existence of a light Higgs boson 
at the LHC or Tevatron is therefore a very important test for both the SM and the MSSM.
Among the various production channels for a light Higgs boson at the hadron colliders,
the production in association with a bottom quark pair, $pp({\rm or~} p\bar p)\to hb\bar b+X$,
plays an important role in testing the bottom quark Yukawa couplings. 
While this process has a  small cross section in the SM, in 
the MSSM this production mechanism can be a significant source of 
Higgs bosons since the bottom quark Yukawa coupling in the MSSM is 
proportional to $\tan\beta$ (defined as $v_2/v_1$ with $v_{1,2}$ being
the vacuum expectation values of the two Higgs doublets) and the current analyses favor  
large $\tan\beta$. This process has been studied at next-to-leading order in perturbative QCD 
~\cite{Dicus,Balazs,Campbell,Dittmaier,Dawson1,Dawson2,Dicus2}. 
Due to the importance of this production mechanism
in testing the bottom quark Yukawa coupling in the MSSM, the supersymmetry (SUSY)
loop effects in this process should also be considered. Of course, among the SUSY
loop effects the one-loop SUSY QCD corrections are the dominant.  
 
In the present work we examine the one-loop SUSY QCD effects in this process. Instead of
performing a complete one-loop calculation, which would be quite complicated since 
it involves many five-point box diagrams, we focus on the so-called SUSY residual 
effects, i.e., the SUSY effects in the heavy limit ($\gsim$ TeV) of the 
sparticles involved. Our motivations are the following:
\begin{itemize}
\item It is possible that all the sparticles except a light Higgs boson are too heavy to be 
      directly produced at the LHC and Tevatron, such as in the split SUSY scenario 
      proposed recently by
      Arkani-Hamed and Dimopoulos \cite{split}. Although the fermionic partners 
      (gauginos and Higgsinos) in this scenario are required to be relatively light in order 
      to ensure gauge coupling unification and provide dark matter candidates, 
      they are not necessarily below a TeV, as recently shown \cite{split2}.
      Thus, it is possible for the LHC and Tevatron to observe no sparticles except a light Higgs 
      boson. In that case a possible way to reveal a hint of supersymmetry is through 
      its residual effects in observable processes.
\item Unfortunately (or fortunately), SUSY virtual effects decouple in most processes 
      when SUSY particles become very heavy. However, we know that the only processes 
      where SUSY has residual effects are those processes involving Higgs-fermion Yukawa 
      couplings, as first studied for $h\to b\bar b$ decay 
      \cite{haber-nondecoup,hbb-loop}, and also for certain production 
      processes \cite{bgbh,bgth}.  
      The production reaction $pp({\rm or~} p\bar p)\to hb\bar b+X$ at the LHC or Tevatron 
      is well suited for revealing 
      SUSY residual effects since it involves the $hb\bar b$ coupling. 
      Compared with  $pp({\rm or~} p\bar p)\to h b+X$, this process is also easier
      to detect since it contains
      an extra hard $b$ jet in the final state. Once the Higgs boson ($h$) is observed 
      and its mass is measured through other processes such as gluon-gluon fusion, this 
      reaction can be used to measure the bottom quark Yukawa coupling and to observe the expected 
      residual effects of SUSY.           
\end{itemize}
Note that the existence of SUSY residual effects in some Higgs process does not mean SUSY 
is not decoupling in low energy processes. As shown in previous studies, and also in the
following work, the residual effects exist when $M_A$ remains light; when $M_A$ is 
heavy, together with all other SUSY masses, the residual effects do vanish.

This paper is organized as follows: In Section \ref{sec:calculation}
we present our strategy for the calculation of the one-loop SUSY QCD 
corrections. In Section \ref{sec:results} we perform numerical calculations
and obtain the residual effects in the limit of heavy SUSY masses. 
The conclusion is given in Section \ref{sec:conclusion} and the 
detailed analytic formulas obtained in our calculations are presented 
in the Appendix.

\section{One-loop SUSY QCD corrections}
\label{sec:calculation}

\begin{figure}[hbt]
\begin{center}
\epsfig{file=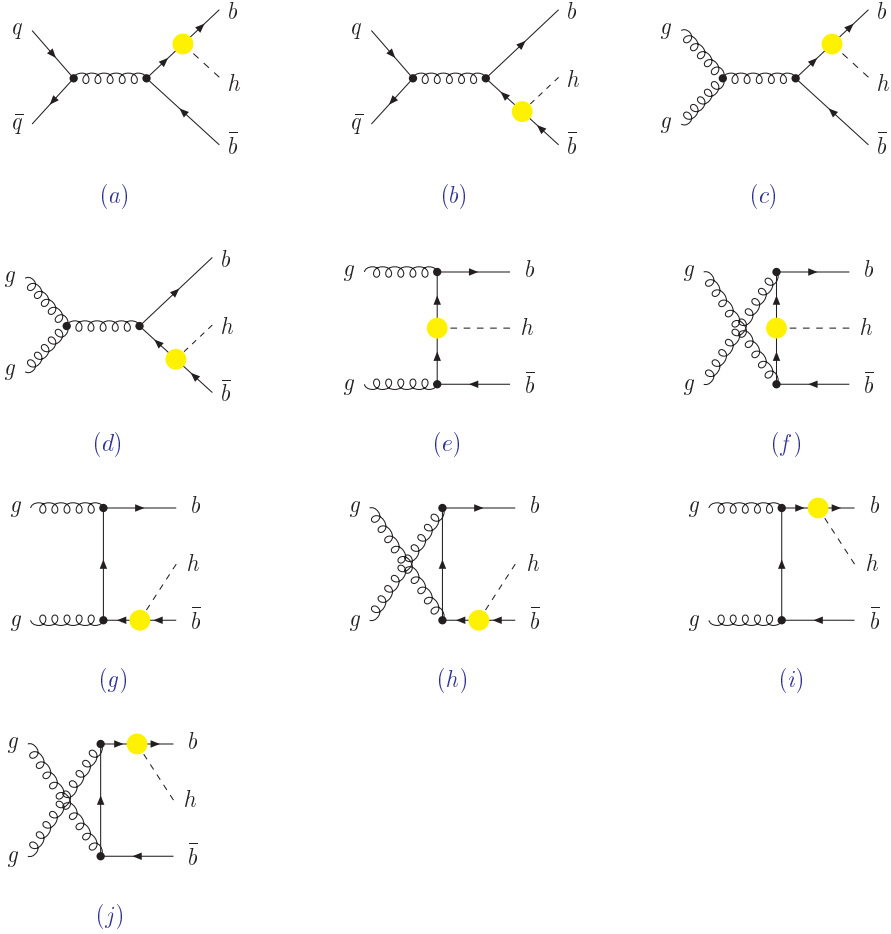,width=13cm}
\caption{\it Feynman diagrams for parton-level subprocesses for 
 $pp({\rm or~} p\bar p)\to hb\bar b+X$ with one-loop SUSY QCD corrected $hb\bar b$ vertices.
The large dots denote the effective $hb\bar b$ vertex with  one-loop SUSY QCD corrections.}
\label{fig1}
\end{center}
\end{figure}

\begin{figure}[hbt]
\begin{center}
\epsfig{file=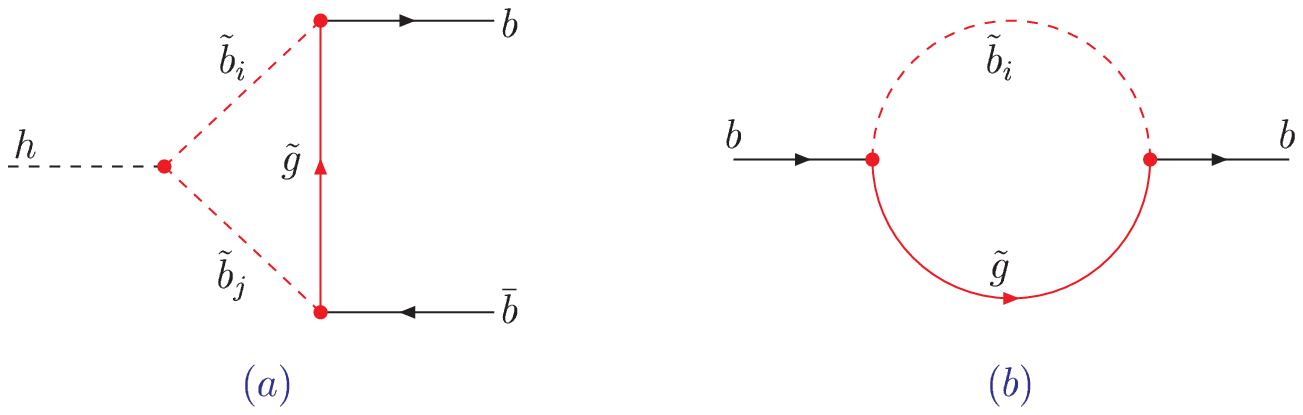,width=13cm}  \vspace*{0.5cm}
\caption{\it Feynman diagrams for the one-loop SUSY QCD corrections to the $hb\bar b$ vertex:
(a) the irreducible vertex loops and (b)the self-energy loops of the $b$ quark.}
\label{fig2}
\end{center}
\end{figure}

The production $pp({\rm or~} p\bar p)\to h b\bar b+X $ proceeds through the 
parton-level processes $gg \to h b\bar b$ and $q \bar q \to h b\bar{b}$.
The one-loop SUSY QCD corrections to this process have a huge number 
of one-loop diagrams, including many five-point box diagrams.
However, among all these diagrams only the one-loop diagrams involving the
bottom quark Yukawa coupling have residual effects as the SUSY masses become 
very heavy. Therefore, in our calculation we need only consider the loop corrected 
bottom quark Yukawa coupling diagrams shown in Fig.~\ref{fig1} and  Fig.~\ref{fig2}.


In our loop calculations we used dimensional regularization to control 
the ultraviolet divergences and adopted the on-mass-shell renormalization scheme.
Each effective $hb\bar b$ vertex in Fig.\ref{fig1} contains two parts: one is 
the irreducible three-point vertex loop contributions and the other is the counterterms 
$\delta V_{hb\bar b}=g_{hb\bar b}^0 \delta Z$ with $g_{hb\bar b}^0$ denoting the
tree-level $hb\bar b$ coupling and $\delta Z$ is the renormalization constant
given by
\begin{equation}
\delta{Z}=\frac{\delta{Z}_L}{2}+\frac{\delta{Z}_R}{2}+\frac{\delta{m}_b}{m_b} \  ,
\end{equation}
where $\delta{Z}_{L,R}$ and $\delta{m}_b$ are respectively the renormalization
constant for the $b$ quark field and mass. They can be extracted from the
one-loop self-energies shown in Fig.~\ref{fig2}(b) by using the on-mass-shell 
renormalization condition.
They are given by
\begin{eqnarray}
\delta Z_L &=& \sum_{i=1}^2\left[ \left( 2m_b^2A_i^1\frac{\partial B_1}{\partial p_b^2}
   -2m_bm_{\tilde{g}}A_i^2\frac{\partial B_0}{\partial p_b^2} \right ) 
   \left \vert_{p_b^2=m_b^2}+(A_i^1-A_i^3)B_1 \right]
   \left(m_b^2, m_{\tilde{g}}^2, m_{\tilde{b}_i}^2 \right) \right. \ ,\\
\delta Z_R &=& \delta Z_L\left \vert_{A_{i}^3\rightarrow -A_{i}^3} \ ,  \right. \\
\frac{\delta m_b}{m_b}&=& \left(\frac{m_{\tilde{g}}}{m_b} A_{i}^2B_0-A_i^1B_1\right)
                  \left(m_b^2, m_{\tilde{g}}^2, m_{\tilde{b}_i}^2\right) \ ,
\end{eqnarray}
where $A_i^1=a_i^2+b_i^2$, $A_i^2=a_i^2-b_i^2$ and $A_i^3=2a_ib_i$ with $a_i$ and  
$b_i$ given in the Appendix. $B_{0,1}$ are the 2-point Feynman integrals given
in \cite{loop}, and their functional dependence is indicated in the brackets following them. 
 
The counterterm is universal for each $hb\bar b$ vertex shown in  Fig.~\ref{fig1}.
However, although the irreducible three-point vertex loops have the same topological structure
shown in Fig.~\ref{fig2}(a), the results are different for different $hb\bar b$ vertices in 
Fig.~\ref{fig1} because they depend on the external momenta.
The results are lengthy and are presented in the Appendix.  
We have checked that all the ultraviolet divergences do
cancel as a result of renormalizability of the MSSM.    

Note that
in our calculations we adopted the so-called on-mass-shell
scheme, in which the renormalized mass $m_b=m_b^0-\delta{m}_b$ ($m_b^0 $ is 
the bare mass) is the physical mass, i.e., the pole of the $b$-quark propagator
\cite{haber-nondecoup,Brat,carena}. The main difference of this scheme with the
$\overline{MS}$ scheme is that in the $\overline{MS}$ scheme a running b-quark 
mass is introduced to absorb the leading part of the corrections (for example, the large 
logarithms in QCD corrections) \cite{Brat}. In calculating SUSY-QCD corrections,
there are no such large logarithms and the on-mass-shell scheme is usually adopted (    
an extensive discussion about this issue was provided in \cite{carena})\footnote{
If we use the $\overline{MS}$ scheme in calculating  SUSY-QCD corrections
to the $hb\bar b$ coupling, i.e., define a running b-quark mass 
to absorb some SUSY-QCD correction effects, we will obtain the approximately same result 
for the ratio of the cross sections with and without SUSY-QCD corrections (
because by doing this we merely relocated some correction effects 
and the total one-loop SUSY-QCD correction effects are not changed).}.

Including the one-loop SUSY-QCD corrections to the bottom quark Yukawa coupling, 
the renormalized amplitude for $pp({\rm or~} p\bar p)\to h b\bar{b}+X$ can be written as
\begin{eqnarray}
M = M_{0}^{q\bar{q}}+\delta M^{q\bar{q}}+M_{0}^{gg}+\delta M^{gg} \ ,
\end{eqnarray}
where $M_{0}$ and $\delta M$ represent the tree-level amplitude 
and one-loop SUSY-QCD corrections, respectively. The detailed expression
for $\delta M$ is given in the Appendix.

In our calculation we performed Monte Carlo integration to obtain the 
hadronic cross section by using the CTEQ5L parton
distribution functions~\cite{pdf} with $Q=m_h$ and requiring the transverse 
momentum of the two b-jets to be larger than 15 GeV. 

To exhibit the size of the corrections we define the ratio 
\begin{equation}
\Delta_{SQCD}=\frac{\sigma-\sigma_0}{\sigma_0} \ ,
\end{equation}
where $\sigma_0$ is the tree-level cross section.

\section{Numerical results}
\label{sec:results}

Before performing numerical calculations, we must choose the
parameters involved. For the SM parameters we used
$m_W=80.448$ GeV, $m_Z=91.187$ GeV, $m_t=178$ GeV, $m_b=4.5$ GeV,
$\sin^2 \theta_W =0.223$, and the two-loop running coupling constant $\alpha_s(Q)$.
For the SUSY parameters, apart from the charged Higgs mass, gluino mass
and $\tan\beta$, the mass parameters of sbottoms are involved.
The mass-squared matrix for the sbottoms takes the form~\cite{susyint}
\begin{equation}
M_{\tilde b}^2 =\left(\begin{array}{cc}
m_{{\tilde b}_L}^2& m_bX_b^\dag\\
 m_bX_b& m_{{\tilde b}_R}^2 \end{array} \right) \ ,
\end{equation}
where
\begin{eqnarray}
m_{{\tilde b}_L}^2 &=& m_{\tilde Q}^2+m_b^2-m_Z^2(\frac{1}{2}
-\frac{1}{3}\sin^2\theta_W)\cos(2\beta) \ , \\
m_{{\tilde b}_R}^2 &=& m_{\tilde D}^2+m_b^2 -\frac{1}{3}m_Z^2 \sin^2\theta_W\cos(2\beta) \ , \\
X_b&=& A_b-\mu\tan\beta \ , 
\end{eqnarray}
Here $m_{\tilde Q}^2$ and $m_{\tilde{D}}^2$ are soft-breaking mass terms 
for the left-handed squark doublet $\tilde Q$ and the right-handed down squark 
$\tilde D$, respectively. $A_b$  is the coefficient of the trilinear term
$H_1 \tilde Q \tilde D$ in the soft-breaking terms and $\mu$ is
the bilinear coupling of the two Higgs doublets in the superpotential.
Thus, the SUSY parameters involved in the sbottom mass matrix are
$m_{\tilde{Q}}$, $m_{\tilde{D}}$, $A_b$, $\mu$ and $\tan\beta$.

The mass-squared matrix is diagonalized by a unitary transformation 
which relates the weak eigenstates $\tilde b_{L,R}$ to the mass eigenstates  
$\tilde b_{1,2}$:
\begin{eqnarray}
\left (\begin{array}{c} \tilde b_1 \\ \tilde b_2 \end{array} \right )
 = \left ( \begin{array}{cc}\cos\theta_b &\sin\theta_b \\
                           -\sin\theta_b &\cos\theta_b \end{array} \right )
\left (\begin{array}{c} \tilde b_L \\ \tilde b_R \end{array} \right )
\end{eqnarray}
with the mixing angle and masses determined by
\begin{eqnarray}
m_{\tilde b_{1,2}}&=&\frac{1}{2}\left[ m_{{\tilde b}_L}^2+m_{{\tilde b}_R}^2
  \mp\sqrt{\left(m_{{\tilde b}_L}^2-m_{{\tilde b}_R}^2\right)^2 +4m_b^2X_b^2}\right] ,\\
\tan2\theta_b &=& \frac{2m_bX_b}{m_{{\tilde b}_L}^2-m_{{\tilde b}_R}^2} \ .
\end{eqnarray}
In addition, we consider the following experimental constraints:
\begin{itemize}
\item[{\rm(1)}]
   $\mu>0$ and large $\tan\beta$, in the range $5\le\tan\beta\le 50$, which are
   favored by the recent muon $g-2$ measurement~\cite{Brown01}.
\item[{\rm(2)}]
   The LEP and CDF lower mass bounds on gluino and sbottom~\cite{PDG00}
\begin{eqnarray}
m_{{\tilde b}_1}\geq 75.0{\rm~GeV},~ m_{\tilde{g}}\geq 190{\rm~GeV} \ .
\end{eqnarray}
\end{itemize}

Next, we present numerical results for LHC 
($\sqrt s = 14$ TeV) and Tevatron ($\sqrt s = 2$ TeV) in two representative cases:
\begin{itemize}
\item[ ] {\em Case A: ~~ All SUSY mass parameters are of the same size and much heavier 
                than the weak scale, but $M_A$ is fixed at weak scale,} i.e.,
\begin{eqnarray}
M_{SUSY} \equiv M_{\tilde Q} = M_{\tilde D}=A_b=M_{\tilde g}=\mu \gg M_{EW} \ .
\label{casea}
\end{eqnarray}
\vspace*{-0.8cm}
\begin{figure}[hbt]
\begin{center}
\epsfig{file=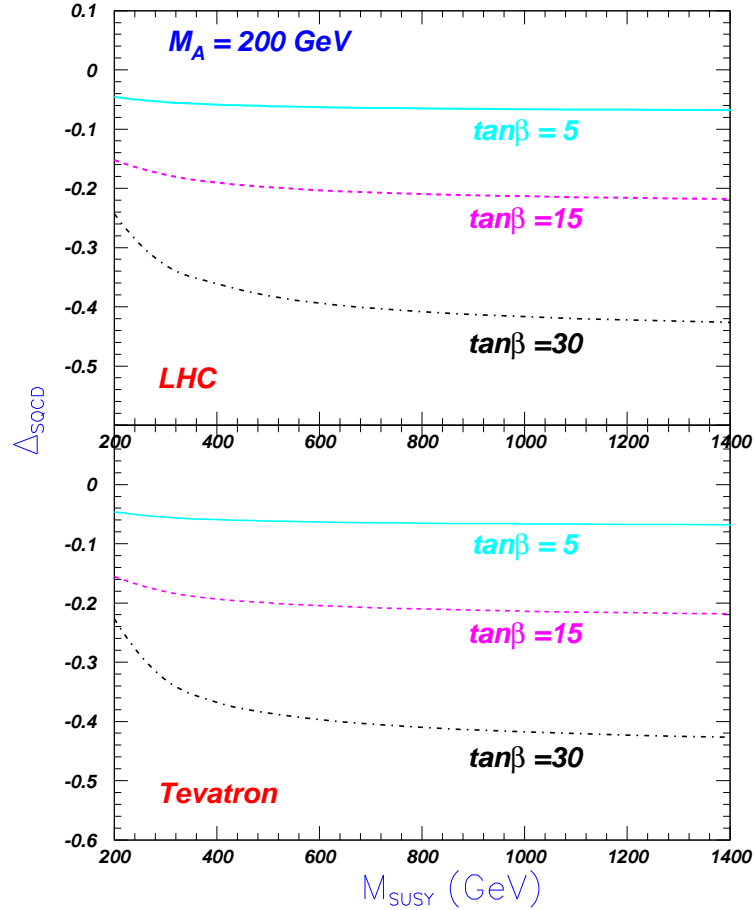,width=10.8cm} \vspace*{-0.2cm}
\caption{\it SUSY QCD effects $\Delta_{SQCD}=(\sigma-\sigma_0)/\sigma_0$ 
         versus the common SUSY mass 
         $M_{SUSY}$($\equiv M_{\tilde Q}=M_{\tilde D}=A_b=M_{\tilde g}=\mu$)
         for a fixed $M_A$ and different values of $\tan\beta$.
         The corresponding range of $m_h$ is $94\sim 113$ GeV for $\tan\beta=5$,  
         $102\sim 121$ GeV for $\tan\beta=15$ and $103\sim 122$ GeV for $\tan\beta=30$.}
\label{fig3}
\end{center}
\end{figure}

\begin{figure}[hbt]
\begin{center}
\epsfig{file=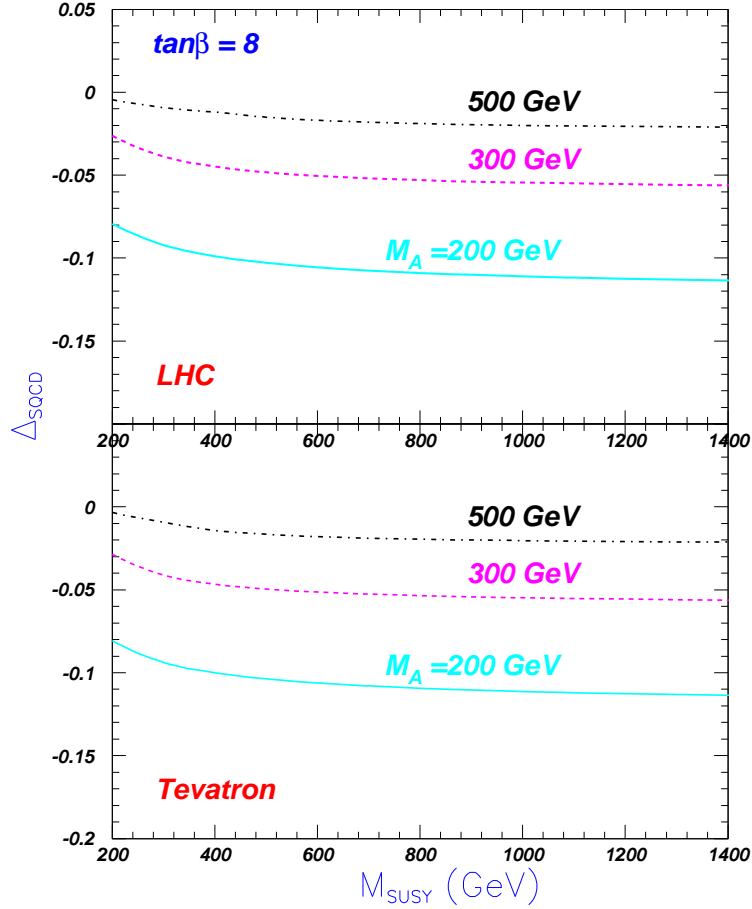,width=10.8cm} \vspace*{-0.2cm}
\caption{\it Same as Fig.\ref{fig3} but for a fixed value of $\tan\beta$ and  
             different values of $M_A$.
             The corresponding range of $m_h$ is $99\sim 118$ GeV for $m_A=200$ GeV, 
             $100\sim 119$ GeV for $m_A=300$ GeV and $100\sim 120$ GeV for $m_A=500$ GeV.}
\label{fig4}
\end{center}
\end{figure}

In this case the mixing of sbottoms is maximal, i.e., $\theta_b\sim\pm \pi/4$. 
Figs.~\ref{fig3} and \ref{fig4}  show the dependence on the SUSY scale $M_{SUSY}$ 
for different values of $\tan\beta$ and $M_A$. 
We see that $\Delta_{SQCD}$ approaches a non-vanishing constant
as $M_{SUSY}$ becomes large. The effects are enhanced by $\tan\beta$. 
For $\tan\beta=30$, the residual effects can be as large as $40\%$ 
\footnote{ Note that when the one-loop effects are very large, higher order loops should
also be considered. In Ref.~\cite{carena} resummation
techniques are proposed to improve the one-loop results.}.
Fig.~\ref{fig4} shows that as $M_A$ becomes large the size of residual effects
decrease.   
\item[] {\em Case B:~~ All SUSY mass parameters, including $M_A$, are much larger
                    than the weak scale,} i.e.,
\begin{eqnarray}
M_{SUSY}\equiv M_{\tilde Q} = M_{\tilde D} = A_b = M_{\tilde g} = \mu = M_A \gg M_{EW} \ .
\end{eqnarray}
This case also gives maximal mixing for sbottoms.
In Fig.~\ref{fig5} we show the dependence on the MSSM mass scale $M_{SUSY}$.
This figure explicitly exhibits the decoupling behaviour of SUSY QCD as $M_{SUSY}$ becomes large.
\end{itemize}

\begin{figure}[hbt]
\begin{center}
\epsfig{file=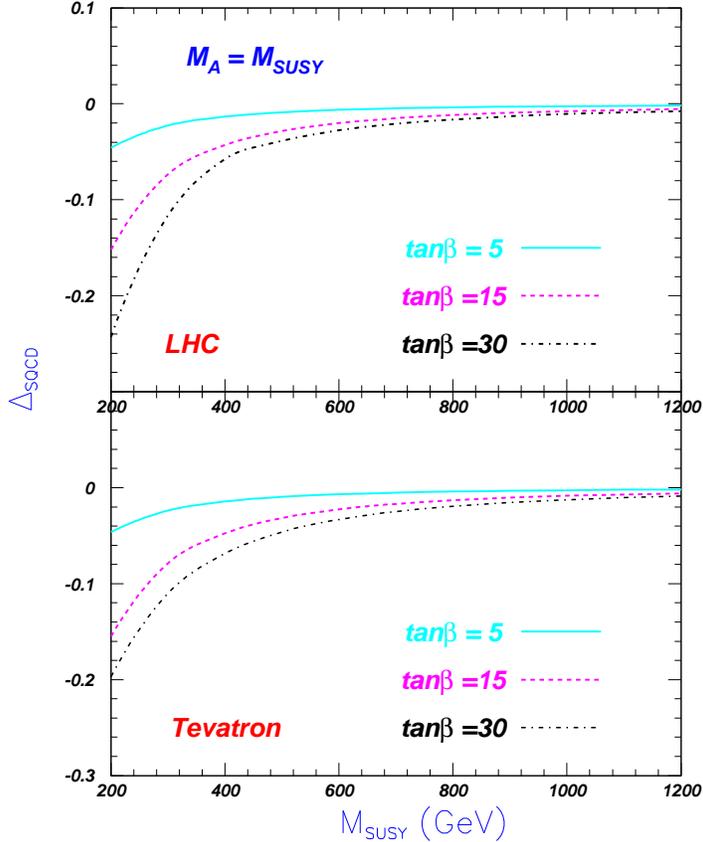,width=10cm} \vspace*{-0.2cm}
\caption{Same as Fig.\ref{fig3} but with $M_A=M_{SUSY}$.
         The corresponding range of $m_h$ is $94\sim 116$ GeV for $\tan\beta=5$,  
         $102\sim 122$ GeV for $\tan\beta=15$ and $103\sim 123$ GeV for $\tan\beta=30$.}
\label{fig5}
\end{center}
\end{figure}

The underlying reason for the existence of SUSY residual effects in fermion Yukawa
couplings when $M_A$ is fixed at the weak scale is that certain couplings are proportional to 
SUSY mass parameters \cite{lixy}. As pointed out in \cite{higgsqq}, 
the SUSY QCD residual effects in the $h b\bar{b}$ coupling are proportional to $\tan\beta$.
For $h t\bar{t}$ coupling, however, the SUSY QCD residual effects are proportional to
$\cot\beta$. For the process $pp({\rm or~}p\bar p)\to h t\bar{t}+X$ the calculation of 
SUSY QCD corrections 
is analogous to the $pp({\rm or~}p\bar p)\to h b\bar{b}+X$ case.
However, we found that the SUSY residual effects in 
$pp({\rm or~}p\bar p)\to h t\bar{t}+X$ are quite small, reaching only $3\%$ for $\tan\beta=5$ and
being smaller for larger $\tan\beta$ values. Such a small effect is even less  
than the next-to-leading-order QCD corrections 
~\cite{Beenakker,Reina,Wbeen,Lreina,Dawsontt,Sdawsontt} and is not likely to
be observed at the LHC or Tevatron.

Note that in our calculations, instead of using  the effective $h b\bar{b}$ vertex \cite{hbb-loop},
we performed the complete one-loop calculations for $h b\bar{b}$ Yukawa coupling, which, as 
pointed out in the paragraph following eq.(4), are dependent on the external momenta of 
$h b\bar{b}$ vertex.
We found that when SUSY mass scale is larger than about 1 TeV, using the effective $h b\bar{b}$ 
vertex \cite{hbb-loop} is a good approximation. 

As we pointed out in the Introduction, we only considered the one-loop SUSY QCD corrections,
which are the dominant part of the SUSY corrections. Of course, the SUSY electroweak
corrections to  $h b\bar{b}$ coupling also have residue effects, which, however, are obviously 
suppressed by a factor $\alpha_{EW}/\alpha_s$ relative to the SUSY QCD effects. Since much more
parameters are involved in the SUSY electroweak sector, we did not perform a detailed 
calculation for the SUSY electroweak corrections here.  

\section{Conclusion}
\label{sec:conclusion}
We examined the supersymmetric QCD effects in bottom pair production 
associated with a light Higgs boson at the LHC and Tevatron.
We found that when the relevant sparticles are heavy, well above the TeV scale, 
they nevertheless contribute sizable virtual effects in this process.
For large $\tan\beta$, these residual effects can alter the production cross section by 
as much as 40 percent and thus should be measurable in observations of this process
at the LHC or Tevatron.  

If only one light Higgs boson, or perhaps several Higgs bosons, are discovered at the LHC
or Tevatron, 
it could indicate that the SUSY scale is quite high; above the TeV scale. Of course, the 
fine-tuning problem then remains and supersymmetry loses one of its merits. But this might 
possibly happen since supersymmetry does not have to solve the fine-tuning problem, as argued 
recently by Arkani-Hamed and Dimopoulos \cite{split}. In that case, a possible way to reveal 
a hint of supersymmetry is through its residue effects. Higgs production associated with
a pair of bottom quarks is then well suited for such a seeking a clue for supersymmetry.

\section*{Acknowledgment}
We thank Junjie Cao for discussions.
This work is supported in part by the Chinese Natural Science Foundation
and by the US Department of Energy, Division of High Energy Physics
under grant No. DE-FG02-91-ER4086.

\appendix
\section{Expressions of SUSY QCD corrections to the amplitude }

The coupling at the vertex $h^0\tilde{b}_{i}\tilde{b}_{j}^*$ is needed in our calculations.
It is given by
\begin{eqnarray}
V(h^0\tilde{b}_{i}\tilde{b}_{j}^*)= i g Q_{ij} \ ,
\end{eqnarray}
where  
\begin{eqnarray}
Q_{11}&=&c_1 \cos^2\theta_b +c_2 \sin^2 \theta_b +2 c_3 \sin\theta_b \cos \theta_b \ ,  \\
Q_{12}&=&(c_2-c_1) \sin{\theta_b}\cos{\theta_b} +c_3 (\cos^2\theta_b -\sin^2\theta_b) \ , \\
Q_{21}&=&(c_2-c_1) \sin{\theta_b}\cos{\theta_b} + c_3(\cos^2\theta_b -\sin^2\theta_b) \ ,  \\
Q_{22}&= &c_1 \sin^2\theta_b +c_2 \cos^2 \theta_b - 2 c_3 \sin{\theta_b}\cos{\theta_b}\ ,
\end{eqnarray}
with
\begin{eqnarray}
c_1 &=& -\frac{m_Z}{\cos{\theta_W}} \left (\frac{1}{2}-\frac{1}{3} \sin^2{\theta_W} \right ) 
                \sin(\alpha+\beta) \ , \\
c_2 &=& -\frac{m_Z}{\cos{\theta_W}} \frac{1}{3} \sin^2{\theta_W} \sin(\alpha+\beta) \ , \\
c_3 &=& \frac{m_b}{2m_W\cos{\beta}}(A_b \sin{\alpha}+\mu \cos{\alpha})\ .
\end{eqnarray}
Here $\alpha$ is the mixing angle between the 
two neutral CP-even Higgs bosons.
Then one-loop contributions $\delta{M}^{q\bar{q}}$ and $\delta{M}^{gg}$ are given by
\begin{eqnarray}
\delta{M}_{1}^{q\bar{q}}&=&\frac{-iG_ST_{ij}^{a}T_{kl}^{a}}{\hat{s}[(k_2+k_3)^2-m_b^2]}
                             \overline{v}(p_1)\gamma^{\mu}u(p_{2})\overline{u}(k_{2})
                  \left [(a_i-b_i\gamma^5)(-\not{k_2}C_{11} 
                  -\not{k_3}C_{12} +m_{\tilde g}C_0) 
                  \right. \nonumber\\
&&\left.\times (a_j+b_j\gamma^5)Q_{ij} -G\delta{Z}\right ] (\not{k_2}+\not{k_3}+m_b)  
 \gamma_{\mu}v(k_{1})(-k_2,-k_3,m_{\tilde g}, m_{\tilde{b}_i},m_{\tilde{b}_j}) \ , \\
\delta{M}_{2}^{q\bar{q}}&=&\frac{iG_ST_{ij}^{a}T_{kl}^{a}}{\hat{s}[(k_1+k_3)^2-m_b^2]}
   \overline{v}(p_1)\gamma^{\mu}u(p_{2})\overline{u}(k_{2})
   \gamma_{\mu}(\not{k_1}+\not{k_3}-m_b)  [(a_i-b_i\gamma^5)(\not{k_1}C_{11}
   \nonumber\\
&& +\not{k_3}C_{12}  +m_{\tilde g}C_0)(a_j+b_j\gamma^5)Q_{ij}
   -G\delta{Z}]v(k_{1})(k_1,k_3,m_{\tilde g}, m_{\tilde{b}_j},m_{\tilde{b}_i}) \ , \\
\delta{M}_{1}^{gg}&=&\frac{-G_ST_{ij}^{c}f_{abc}} {\hat{s}[(k_2+k_3)^2-m_b^2]}\overline{u}(k_{2})
                   [(a_i-b_i\gamma^5)(-\not{k_2}C_{11} -\not{k_3}C_{12} 
                    +m_{\tilde g}C_0)(a_j+b_j\gamma^5)Q_{ij}  \nonumber\\
&&-G\delta{Z}]
    (\not{k_2}+\not{k_3}+m_b)\gamma^{\lambda}v(k_{1})
   F_{\lambda\mu\nu}\varepsilon^{\mu}(p_1) \varepsilon^{\nu}(p_2)(-k_2,-k_3,m_{\tilde g},
   m_{\tilde{b}_i},m_{\tilde{b}_j})\ , \\
\delta{M}_{2}^{gg}&=&\frac{G_ST_{ij}^{c}f_{abc}}{\hat{s}[(k_2+k_3)^2-m_b^2]}
             \overline{u}(k_{2})\gamma^{\lambda}(\not{k_1}+\not{k_3}-m_b)
             [(a_i-b_i\gamma^5) 
    (\not{k_1}C_{11}+\not{k_2}C_{12}+m_{\tilde g}C_0)
   \nonumber\\
&&\times  (a_j+b_j\gamma^5)Q_{ij}-G\delta{Z}]v(k_{1})
   F_{\lambda\mu\nu}\varepsilon^{\mu}(p_1) \varepsilon^{\nu}(p_2)(k_1,k_3,m_{\tilde g},
   m_{\tilde{b}_j},m_{\tilde{b}_i}) \ , \\
\delta{M}_{3}^{gg}&=&\frac{iG_ST_{ij}^{b}T_{jk}^{a}}{[(k_2+p_2)^2-m_b^2)][(p_1-k_1)-m_b^2]}
                \overline{u}(k_{2})\gamma_{\nu}(\not{k_2}-\not{p_2} 
             +m_b)[(a_i-b_i\gamma^5)(\not{k_1}C_{11}
	       \nonumber\\
&& -\not{p_1}C_{11}
    +\not{k_3}C_{12}+m_{\tilde g}C_0) 
   (a_j+b_j\gamma^5)Q_{ij}-G\delta{Z}]\nonumber\\
&& \times (\not{p_1}-\not{k_1}+m_b)\gamma_{\mu}v(k_{1})
   \varepsilon^{\mu}(p_1)\varepsilon^{\nu}(p_2)(k_1-p_1,k_3,m_{\tilde g},
    m_{\tilde{b}_j},m_{\tilde{b}_i}) \ , \\
\delta{M}_{4}^{gg}&=&\frac{iG_ST_{ij}^{b}T_{jk}^{a}}{[(k_2+k_3)^2-m_b^2)][(p_1-k_1)-m_b^2]}
   \overline{u}(k_{2})[(a_i-b_i\gamma^5) (-\not{k_2}C_{11}-\not{k_3}C_{12} \nonumber\\
&& +m_{\tilde g}C_0) (a_j+b_j\gamma^5)Q_{ij}-G\delta{Z}] (\not{k_2}+\not{k_3}+m_b)\gamma_{\nu}
   \nonumber\\
&& \times (\not{p_1}-\not{k_1}+m_b)\gamma_{\mu}v(k_{1}) 
  \varepsilon^{\mu}(p_1)\varepsilon^{\nu} (p_2)(-k_2,-k_3,m_{\tilde g},
   m_{\tilde{b}_i},m_{\tilde{b}_j}) \ , \\
\delta{M}_{5}^{gg}&=&\frac{iG_ST_{ij}^{a}T_{jk}^{b}}
    {[(k_2+p_1)^2-m_b^2)][(p_2-k_1)-m_b^2]}\overline{u}
    (k_{2})\gamma_{\mu}(\not{k_2}-\not{p_1} +m_b)[(a_i-b_i\gamma^5)(\not{k_1}C_{11}
    \nonumber\\
&& -\not{p_2}C_{11}
   +\not{k_3}C_{12}+m_{\tilde g}C_0)  (a_j+b_j\gamma^5)Q_{ij}
   -G\delta{Z}](\not{p_2} -\not{k_1} 
   \nonumber\\
&&+m_b)\gamma_{\nu}v(k_{1})
  \varepsilon^{\mu}(p_1)\varepsilon^{\nu}(p_2)
   (k_1-p_2,k_3,m_{\tilde g}, m_{\tilde{b}_j},m_{\tilde{b}_i}) \ , \\
\delta{M}_{6}^{gg}&=&\frac{iG_ST_{ij}^{a}T_{jk}^{b}}{[(k_2+k_3)^2-m_b^2)][(p_2-k_1)-m_b^2]}
                  \overline{u}(k_{2})[(a_i-b_i\gamma^5) 
 (-\not{k_2}C_{11}-\not{k_3}C_{12}+m_{\tilde g}C_0)
  \nonumber\\
&&\times (a_j+b_j\gamma^5)Q_{ij}-G\delta{Z}]  (\not{k_2}+\not{k_3}+m_b)\gamma_{\mu}
   (\not{p_2}-\not{k_1} \nonumber\\
&& +m_b)\gamma_{\nu}v(k_{1})
  \varepsilon^{\mu}(p_1)\varepsilon^{\nu}(p_2)(-k_2,-k_3,m_{\tilde g},
   m_{\tilde{b}_i},m_{\tilde{b}_j}) \ , \\
\delta{M}_{7}^{gg}&=&\frac{-iG_ST_{ij}^{a}T_{jk}^{b}}{[(k_2-P_1)^2-m_b^2)][(k_1+k_3)-m_b^2]}
             \overline{u}(k_{2})\gamma_{\mu}(\not{k_2}-\not{p_1}
  +m_b)\gamma_{\nu}(\not{k_1}+\not{k_3} \nonumber\\
&&-m_b) [(a_i-b_i\gamma^5)(\not{k_1}C_{11}+\not{k_3}C_{12}
  +m_{\tilde g}C_0)(a_j+b_j\gamma^5)Q_{ij}
 \nonumber\\
&&
-G\delta{Z}]v(k_{1}) \varepsilon^{\mu}(p_1)\varepsilon^{\nu}(p_2)(k_1,k_3,m_{\tilde g},
   m_{\tilde{b}_j},m_{\tilde{b}_i}) \ , \\
\delta{M}_{8}^{gg}&=&\frac{-iG_ST_{ij}^{b}T_{jk}^{a}}{[(k_2-P_2)^2-m_b^2)][(k_1+k_3)-m_b^2]}
   \overline{u}(k_{2})\gamma_{\nu}(\not{k_2}-\not{p_2}
   +m_b)\gamma_{\mu}(\not{k_1}+\not{k_3} \nonumber\\
&&-m_b)[(a_i-b_i\gamma^5)(\not{k_1}C_{11}+\not{k_3}C_{12} 
   +m_{\tilde g}C_0)(a_j+b_j\gamma^5)Q_{ij}
  \nonumber\\
&&
-G\delta{Z}]v(k_{1}) \varepsilon^{\mu}(p_1)\varepsilon^{\nu}(p_2)(k_1,k_3,m_{\tilde g},
   m_{\tilde{b}_j},m_{\tilde{b}_i}) \ ,
\end{eqnarray}
where we defined
\begin{eqnarray}
a_{1,2}&=&(\sin{\theta_b}\mp\cos{\theta_b})/\sqrt 2 \ , \\ 
b_{1,2}&=&(\cos{\theta_b}\pm\sin{\theta_b})/\sqrt 2 \ , \\
F_{\lambda\mu\nu}&=&(p_1-p_2)_{\lambda}g_{\mu\nu} +(p_2+k_1+k_2+k_3)_{\mu}g_{\nu\lambda}
                   -(p_1+k_1+k_2+k_3)_{\nu}g_{\lambda\mu} \ .
\end{eqnarray}
In the above, $p_1$ and $p_2$ are respectively the momenta of the incoming quark and antiquark,
$k_1$,$k_2$ and $k_3$ are respectively the outgoing $b$ quark, $\bar b$ quark, and the 
Higgs boson momenta, $\hat{s}$ is the Mandelstam variable defined by 
$\hat{s}=(k_{1}+k_{2}+k_{3})^2=(p_1+p_2)^2$, $T^{a}$ are the $SU(3)$ color matrices,
$f_{abc}$ are anti-symmetric structure constants of $SU(3)$ color matrices, 
$G_S=gg_s^4C_F/16\pi^2$ with $C_F=4/3$,  and $C_0$ and $C_{ij}$ are the 3-point Feynman integrals 
\cite{loop} with their functional dependence indicated in the brackets following them.

\end{document}